\documentclass[12pt]{article}

\usepackage{amsmath,amssymb,bbm,graphicx}

\usepackage{float}
\newfloat{figure}{H}{lof}
\newfloat{table}{H}{lot}
\floatname{figure}{\figurename}
\floatname{table}{\tablename}
\floatname{label}{\labelname}

\newcommand{\assign}{:=}
\newcommand{\email}[1]{{\textit{Email:} \texttt{#1}}}

\newcommand{\tmem}[1]{{\em #1\/}}
\newcommand{\tmmathbf}[1]{\ensuremath{\boldsymbol{#1}}}
\newcommand{\tmop}[1]{\ensuremath{\operatorname{#1}}}
\newcommand{\tmstrong}[1]{\textbf{#1}}
\newcommand{\tmtextbf}[1]{{\bfseries{#1}}}
\newcommand{\tmtextit}[1]{{\itshape{#1}}}
\newcommand{\tmtextup}[1]{{\upshape{#1}}}

\begin{document}

\date{}

\title{On the mean Euler characteristic and mean Betti's numbers of the
Ising model with arbitrary spin\footnote{Preprint CPT--P01--2006; This document has been written using the GNU TeXmacs text editor (see www.texmacs.org)}} \author{Philippe
BLANCHARD\thanks{\email{blanchard@physik.uni-bielefeld.de}}\\
Fakult\"at f\"ur Physik, Theoretishe Physik and Bibos, Universit\"at
Bielefeld,\\ Postfach 100131, D--33501, Bielefeld, Germany. \and
Christophe DOBROVOLNY\thanks{\email{christophe.dobrovolny@ucd.ie}}\\
University College Dublin, Belfield, Dublin 4, Ireland. \and Daniel
GANDOLFO\thanks{\email{gandolfo@cpt.univ-mrs.fr}}\\
Centre de Physique Th\'eorique, CNRS Luminy case 907, F--13288\\ Marseille
Cedex 9, France, and D\'epartement de Math\'ematiques,\\ Universit\'e du Sud
Toulon--Var, BP 132, F--83957 La Garde, France. \and Jean
RUIZ\thanks{\email{ruiz@cpt.univ-mrs.fr}}\\
Centre de Physique Th\'eorique, CNRS Luminy case 907,\\ F--13288 Marseille
Cedex 9, France.} 

\maketitle

\begin{abstract}
  The behaviour of the mean Euler--Poincar\'e characteristic and mean Betti's
  numbers in the Ising model with arbitrary spin on $\mathbbm{Z}^2$ as functions of the temperature is investigated through intensive
  Monte Carlo simulations. We also consider these quantities for each color $a$  in the state space $S_Q = \{ - Q, - Q + 2, \ldots, Q \}$ of the model.
  We find that these topological invariants show a sharp transition at the
  critical point.
\end{abstract}


\section{Introduction}

Knowledge about the spatial structure of systems is a subject that attracts
more and more interest in statistical physics. For example, the study of
morphological features has proved very useful to identify and distinguish
various formation processes in contexts as varied as the study of large scale
distributions of galaxies, or the investigation of micro-emulsions
{\cite{Ser}}. Other examples of situations where the focus on the spatial
structure is of primary importance to understand the physical properties of a
given system are provided by dissipative structures in hydrodynamics, Turing
patterns occurring in chemical reactions, or transport properties of fluids
{\cite{Me1}}. For instance, the transport of a fluid over a porous substrate
depends crucially on the existence of a connected cluster of pores that spans
through the whole system {\cite{Me2}}. Gaining insight into the spatial
structure of the most commonly used lattice-models in statistical physics
-i.e. the Potts, Clock and generalized Ising models- is also a matter of interest in
the field of image-analysis, which makes extensive use of these models to
simulate noise and clean dirty images {\cite{Win}}. In this context,
mathematical quantities that characterize the typical spatial structure of the
model seem to be the most natural tuning parameters as opposed to the usual
thermodynamic quantities.\\
{\noindent}Consequently, efforts have recently been made {\cite{Me2,Wa,BGRS}}
to study the typical behavior of Minkowsky-functionals (or curvature
integrals) which quantify the geometrical properties of a given system.
Particular interest has been devoted to the study of the Euler-Poincar\'e
characteristic which turns out to be a relevant quantity in various models :
let us indeed recall, that for the problem of bond percolation on regular
lattices, Sykes and Essam {\cite{SyEs}} were able to show, using standard
planarity arguments, that for the case of self-dual lattices (e.g.
$\mathbb{Z}^2$), the mean value of the Euler-Poincar\'e characteristic changes
sign at the critical point, see also {\cite{Gri}}.\\
{\noindent}More recently Blanchard, {\tmem{et al.}} showed {\cite{BGRS}}
using duality and perturbative arguments, that for the two dimensional random
cluster model (the Fortuin-Kasteleyn representation of the q-state Potts
model) the mean local Euler-Poincar\'e characteristic is either zero or
exhibits a jump at the self-dual temperature. More precisely, it vanishes when
$q = 1$ and exhibits a jump of order $1 / \sqrt{q}$ when $q$ is large enough.
Similar results were shown in higher dimensions $d$.\\
{\noindent}This paper aims to investigate the spatial structure of the
Ising model with arbitrary spins on $\mathbb{Z}^2$ which, unlike the Potts-model, does not have
a color-symmetry. We introduce colored complexes (i.e. the sets of sites for a
given configuration that belong to the same color) and study the behavior of
the associated local mean Euler-Poincar\'e characteristics. As a result of our
numerical computations, we obtain that the mean value of the Euler
characteristic per site vanishes below the critical temperature and is
positive above this temperature. As we shall see, this behaviour can be
related to the behaviours of topological invariants, the Betti's numbers.

{\noindent}The paper is organized as follows. In Section 2, we introduce the
model and give the definitions of the principal quantities of interest.
Section 3 is devoted to the results obtained by intensive Monte-Carlo
simulations. Concluding remarks are given in section $4$.\\

\section{Definitions}

To introduce the generalized Ising model considered by Griffiths' in {\cite{Gr1}}, we associate to each lattice
site $x \in \mathbbm{Z}^2$ a spin $\sigma_x$ in the set $S_Q = \{ - Q, - Q +
2, \ldots, Q \}$, $Q = 1, 2, \ldots,$ (with cardinality $|S_Q | = Q + 1$). The
Hamiltonian in a finite box $\Lambda \subset \mathbbm{Z}^2$ is given by:
\begin{equation}
  H_{\Lambda} = - J \sum_{\langle x, y \rangle} \sigma_x \sigma_y
  \label{hamiltonien}
\end{equation}
where $J$ is a positive constant that we will take equal to $1$ in the next
section, and the sum runs over nearest neighbors pairs of $\Lambda$. The
associated partition function is defined by:
\begin{equation}
  Z_{\Lambda} (\beta) = \sum_{{\begin{array}{c}
    \sigma_x \in S_Q\\
    x \in \Lambda
  \end{array}}} e^{- \beta H_{\Lambda}}
\end{equation}
where the sum is over all configurations $\tmmathbf{\sigma}_{\Lambda} = \{
\sigma_x \}_{x \in \Lambda}$ and we let
\begin{equation}
  \langle f \rangle_{\Lambda} (\beta) = \frac{1}{Z_{\Lambda} (\beta)} 
  \sum_{{\begin{array}{c}
    \sigma_x \in S_Q\\
    x \in \Lambda
  \end{array}}} f e^{- \beta H_{\Lambda}}
\end{equation}
denote the (corresponding expectation) of a measurable function $f$. Among the
quantity of interest, we shall first consider the magnetization
\begin{equation}
  m_{\Lambda} (\beta) = \frac{1}{| \Lambda |} \langle \sum_{x \in \Lambda}
  \sigma_x \rangle_{\Lambda} (\beta)
\end{equation}
where $| \Lambda |$ is the number of sites of $\Lambda$. Let us recall that
this system exhibits a spontaneous magnetization at all inverse temperatures
greater than the critical temperature of the Ising model ($Q = 1$), as proved
in {\cite{Gr1}}.

We next introduce others quantities, namely the mean Euler characteristic per
site. Consider a configuration $\tmmathbf{\sigma}_{\Lambda}$, and let $L
(\tmmathbf{\sigma}_{\Lambda})$ be the set of bonds (unit segments) with
endpoints $x, y$, such that $\sigma_x = \sigma_y$, and $P
(\tmmathbf{\sigma}_{\Lambda})$ be the set of plaquettes $p$ with corner $x, y,
z, t$, such that $\sigma_x = \sigma_y = \sigma_z = \sigma_t$. To a given
configuration $\tmmathbf{\sigma}_{\Lambda}$, we associate (in a unique way)
the two--dimensional cell--subcomplex $C (\tmmathbf{\sigma}_{\Lambda}) = \{
\Lambda, L (\tmmathbf{\sigma}_{\Lambda}), P (\tmmathbf{\sigma}_{\Lambda}) \}$
of the complex $C = \{ \Lambda, L_{\Lambda}, P_{\Lambda} \}$ where
$L_{\Lambda}$ is the set of bonds with both endpoints in $\Lambda$ and
$P_{\Lambda}$ is the set of plaquettes with corners in $\Lambda$. The Euler
characteristic of this subcomplex is defined by:
\begin{equation}
  \tmmathbf{\chi}(\tmmathbf{\sigma}_{\Lambda}) = | \Lambda | - |L
  (\tmmathbf{\sigma}_{\Lambda}) | + |P (\tmmathbf{\sigma}_{\Lambda}) |
\end{equation}
where $|E|$ denotes hereafter the cardinality of the set $E$. It satisfies the
Euler--Poincar\'e formula
\begin{equation}
  \tmmathbf{\chi}(\tmmathbf{\sigma}_{\Lambda}) =\tmmathbf{\pi}^0
  (\tmmathbf{\sigma}_{\Lambda}) -\tmmathbf{\pi}^1
  (\tmmathbf{\sigma}_{\Lambda}) +\tmmathbf{\pi}^2
  (\tmmathbf{\sigma}_{\Lambda}) \label{EP}
\end{equation}
where $\tmmathbf{\pi}^0 (\tmmathbf{\sigma}_{\Lambda})$ and $\tmmathbf{\pi}^1
(\tmmathbf{\sigma}_{\Lambda})$ are respectively the {\tmem{number of connected
components}} and the {\tmem{number of independent $1$-cycles}} of the
subcomplex $C (\tmmathbf{\sigma}_{\Lambda})$. Here $\tmmathbf{\pi}^2
(\tmmathbf{\sigma}_{\Lambda}) = 0$ because $C (\tmmathbf{\sigma}_{\Lambda})$
has no $2$-cycles.

Notice that
\begin{equation}
  \tmmathbf{\chi}(\sigma_{\Lambda}) = \sum_{x \in \Lambda} \tmmathbf{\chi}_{x,
  \tmop{loc}} (\tmmathbf{\sigma}_{\Lambda})
\end{equation}
where the local Euler characteristic $\tmmathbf{\chi}_{x, \tmop{loc}}
(\tmmathbf{\sigma}_{\Lambda})$ is given by:
\begin{equation}
  \tmmathbf{\chi}_{x, \tmop{loc}} (\tmmathbf{\sigma}_{\Lambda}) = 1 -
  \frac{1}{2} \sum_{y \in V (x)} \delta (\sigma_x, \sigma_y) + \frac{1}{4}
  \sum_{p \ni x} \prod_{y z \in p} \delta (\sigma_y, \sigma_z) \label{qlocal}
\end{equation}
Here, $V (x)$ is the set of nearest neighbours of $x$, the second sums is over
plaquettes containing $x$ as a corner, and $\delta$ is the kronecker symbol:
$\text{} \delta (\sigma_x, \sigma_y) = 1$ if $\sigma_x = \sigma_y$, and 
$\delta (\sigma_x, \sigma_y) = 0$ otherwise. We define the mean Euler
characteristic per site and the mean Betti's numbers per site by:
\begin{eqnarray*}
  \chi_{\Lambda} (\beta) & = & \frac{1}{| \Lambda |} \langle
  \tmmathbf{\chi}(\tmmathbf{\sigma}_{\Lambda}) \rangle_{\Lambda} (\beta)
  \label{chi} \hspace*{\fill}   \text{\tmtextup{(1)}}\\
  \pi^0_{\Lambda} (\beta) & = & \frac{1}{| \Lambda |} \langle \tmmathbf{\pi}^0
  (\tmmathbf{\sigma}_{\Lambda}) \rangle_{\Lambda} (\beta) \label{pi0}
  \hspace*{\fill}   \text{\tmtextup{(2)}}\\
  \pi^1_{\Lambda} (\beta) & = & \frac{1}{| \Lambda |} \langle \tmmathbf{\pi}^1
  (\tmmathbf{\sigma}_{\Lambda}) \rangle_{\Lambda} (\beta) \label{pi1}
  \hspace*{\fill}   \text{\tmtextup{(3)}}
\end{eqnarray*}

The following limits exist and coincide:
\begin{equation}
  \lim_{\Lambda \uparrow \mathbbm{Z}^2} \chi_{\Lambda} (\beta) = \lim_{\Lambda
  \rightarrow \mathbbm{Z}^2} \langle \tmmathbf{\chi}_{0, \tmop{loc}}
  (\tmmathbf{\sigma}_{\Lambda}) \rangle_{\Lambda} (\beta) \label{lim}
\end{equation}
Here the first limit is taken in such a way that the number of sites of the
boundary of $\Lambda$ divided by the number of sites of $\Lambda$ tends to
$0$, while there is no restriction in the second limit.

The existence of these limits are simple consequences of Griffith's inequality
{\cite{Gr}} which states that:
\begin{equation}
  \langle \sigma_x^n \sigma_y^m \rangle_{\Lambda} (\beta) \geqslant \langle
  \sigma_x^n \rangle_{\Lambda} (\beta) \langle \sigma_y^m \rangle_{\Lambda}
  (\beta)
\end{equation}
This implies the following monotonicity properties with respect to the volume:
\[ \left\langle \sigma_x^n \sigma_y^m \rangle_{\Lambda} (\beta) \leqslant
   \left\langle \sigma_x^n \sigma_y^m \rangle_{\Lambda'} (\beta) \tmop{if}
   \Lambda \subset \Lambda'  \right. \right. \]
These properties give the existence of the infinite volume limit 
$$\langle \sigma_x^n \sigma_y^m \rangle_{\Lambda} (\beta) \assign 
{\lim_{\Lambda \rightarrow \mathbbm{Z}^2}}
\langle \sigma_x^n \sigma_y^m
\rangle_{\Lambda} (\beta)
$$ 
through sequences of increasing volumes. 
They imply also, following {\cite{Gr}}, that the limiting state is
translation invariant. It is then sufficient to observe that:
\[ \delta (\sigma_x, \sigma_y) = \frac{\prod_{b \in S_Q, b \neq 0} [b -
   (\sigma_x - \sigma_y)]}{\prod_{b \in S_Q, b \neq 0} b} \]
and thus  can be written as a (finite) linear combination $\sum_{n, m \in S_Q}
\lambda_{a, b} \sigma_x^n \sigma_y^m$ with finite coefficients $\lambda_{a,
b}$ to conclude the existence of the second limit in (\ref{lim}), since also
by (\ref{qlocal}) the local Euler characteristic is a (finite) linear
combination of the $\sigma_x^n \sigma_y^m$ with finite coefficients. The
translation invariance allows finally to show that this second limit in
(\ref{lim}) coincides with the first one under the above mentioned conditions.

We shall also consider the subcomplexes of $C (\tmmathbf{\sigma}_{\Lambda})$
attached to each color. Namely, for a given configuration
$\tmmathbf{\sigma}_{\Lambda}$ and a given color $a \in S_Q$, we let
$\Lambda_a$ be the subset of the box $\Lambda$ such that $\sigma_x = a$,  $L_a
(\tmmathbf{\sigma}_{\Lambda})$ be the subset of bonds of $L
(\tmmathbf{\sigma}_{\Lambda})$ with endpoints $x, y,$ such that $\sigma_x =
\sigma_y = a$, and $P_a (\tmmathbf{\sigma}_{\Lambda})$ be the subset of
plaquettes of $P (\tmmathbf{\sigma}_{\Lambda})$ with corner $x, y, z, t$, such
that $\sigma_x = \sigma_y = \sigma_z = \sigma_t = a$. For each color $a$, we
associate to a given configuration $\tmmathbf{\sigma}_{\Lambda}$, the
two--dimensional cell--subcomplex $C_a (\tmmathbf{\sigma}_{\Lambda}) = \{
\Lambda_a, L_a (\tmmathbf{\sigma}_{\Lambda}), P_a
(\tmmathbf{\sigma}_{\Lambda}) \}$. Obviously $C_a
(\tmmathbf{\sigma}_{\Lambda})$ is a subcomplex of  $C
(\tmmathbf{\sigma}_{\Lambda})$ and hence a subcomplex of $C$ and $C
(\tmmathbf{\sigma}_{\Lambda}) = \cup_{a \in S_Q} \text{$C_a
(\tmmathbf{\sigma}_{\Lambda})$}$.

The Euler characteristics of these subcomplexes are defined by:
\begin{equation}
  \tmmathbf{\chi}^a (\tmmathbf{\sigma}_{\Lambda}) = | \Lambda_a | - |L_a
  (\tmmathbf{\sigma}_{\Lambda}) | + |P_a (\tmmathbf{\sigma}_{\Lambda}) |
\end{equation}
They satisfy the Euler--Poincar\'e formula
\begin{equation}
  \tmmathbf{\chi}^a (\tmmathbf{\sigma}_{\Lambda}) =\tmmathbf{\pi}^{0, a}
  (\tmmathbf{\sigma}_{\Lambda}) -\tmmathbf{\pi}^{1, a}
  (\tmmathbf{\sigma}_{\Lambda}) +\tmmathbf{\pi}^{2, a}
  (\tmmathbf{\sigma}_{\Lambda}) \label{EPformula}
\end{equation}
where $\tmmathbf{\pi}^{0, a} (\tmmathbf{\sigma}_{\Lambda})$ and
$\tmmathbf{\pi}^{1, a} (\tmmathbf{\sigma}_{\Lambda})$ are respectively the
number of connected components and the maximal number of independent
$1$-cycles of the subcomplex $C_a (\tmmathbf{\sigma}_{\Lambda})$. Here again
$\tmmathbf{\pi}^{2, a} (\tmmathbf{\sigma}_{\Lambda}) = 0$ because $C_a
(\tmmathbf{\sigma}_{\Lambda})$ has no $2$-cycles.

Note that
\begin{equation}
  \tmmathbf{\chi}(\tmmathbf{\sigma}_{\Lambda}) = \sum_{a \in S_Q}
  \tmmathbf{\chi}^a (\tmmathbf{\sigma}_{\Lambda}) \label{chi-additivity}
\end{equation}
As for the total Euler characteristic, the Euler characteristic of each color
can be written as a sum over local variables:
\begin{equation}
  \tmmathbf{\chi}^a (\sigma_{\Lambda}) = \sum_{x \in \Lambda}
  \tmmathbf{\chi}^a_{x, \tmop{loc}} (\tmmathbf{\sigma}_{\Lambda})
\end{equation}
where
\begin{equation}
  \tmmathbf{\chi}^a_{x, \tmop{loc}} (\tmmathbf{\sigma}_{\Lambda}) = \delta
  (\sigma_x, a) - \frac{1}{2} \sum_{y \in V (x)} \delta (\sigma_x, a) \delta
  (\sigma_y, a) + \frac{1}{4} \sum_{p \ni x} \prod_{y z \in p} \delta
  (\sigma_y, a) \delta (\sigma_z, a) \label{qlocalcolor}
\end{equation}

We define the mean Euler characteristic per site and the mean Betti's numbers
per site of the colored subcomplexes:
\begin{eqnarray*}
  \chi^a_{\Lambda} (\beta) & = & \frac{1}{| \Lambda |} \langle
  \tmmathbf{\chi}^a (\sigma_{\Lambda}) \rangle_{\Lambda} (\beta)
  \label{chicol} \hspace*{\fill}   \text{\tmtextup{(4)}}\\
  \pi^{0, a}_{\Lambda} (\beta) & = & \frac{1}{| \Lambda |} \langle
  \text{$\tmmathbf{\pi}^{0, a} (\tmmathbf{\sigma}_{\Lambda})$}
  \rangle_{\Lambda} (\beta) \label{pi0col} \hspace*{\fill}  
  \text{\tmtextup{(5)}}\\
  \pi^{1, a}_{\Lambda} (\beta) & = & \frac{1}{| \Lambda |} \langle
  \text{$\tmmathbf{\pi}^{1, a} (\tmmathbf{\sigma}_{\Lambda})$}
  \rangle_{\Lambda} (\beta) \label{pi1col} \hspace*{\fill}  
  \text{\tmtextup{(6)}}
\end{eqnarray*}

One can show that the following limit exist and coincide:
\begin{equation}
  \lim_{\Lambda \uparrow \mathbbm{Z}^2} \chi^a_{\Lambda} (\beta) =
  \lim_{\Lambda \rightarrow \mathbbm{Z}^2} \langle \chi^a_{0, \tmop{loc}}
  (\tmmathbf{\sigma}_{\Lambda}) \rangle_{\Lambda} (\beta) \label{lima}
\end{equation}
by using the relation
\begin{equation}
  \delta (\sigma_x, a) = \frac{\prod_{b \in S_Q, b \neq a} (\sigma_x -
  b)}{\prod_{b \in S_Q, b \neq a} (a - b)}
\end{equation}
and arguing as before.

\section{Results}

We have performed intensive Monte Carlo simulations to study the thermodynamic
behaviour of this model. As expected, we find that it belongs to the 2D-Ising
universality class for all tested values of $Q$ with the same critical
exponents. The inverse critical temperature $\beta_c$ at which second order
phase transition occurs for $Q$ ranging from 2 upward tends to 0 with
increasing values of $Q$.

Using Binder cumulants {\cite{BIN}} (see also \cite{NB},\cite{CD},\cite{SS}), we find $\beta \overset{}{}_{c
\text{}}^{Q = 1} = 0.4405 \pm 0.0004$, $\beta \overset{}{}_{c \text{}}^{Q = 2}
= 0.1475 \pm 0.0005$, $\beta \overset{}{}_{c \text{}}^{Q = 5} = 0.0316 \pm
0.0005$. The critical exponent for the magnetization is found equal to $0.125
\pm 0.001 \tmop{and} \tmop{the}$ one for the susceptibility to $1.76 \pm
0.002$.

Figs. 1 and 2 show the finite size scaling plots of the magnetization and the
susceptibility for $Q = 2$ and $Q = 5$ for different system sizes $L$: $|
\Lambda | = L \times L$. 

\begin{figure}[H]
   \begin{center}
    \includegraphics[width=12truecm]{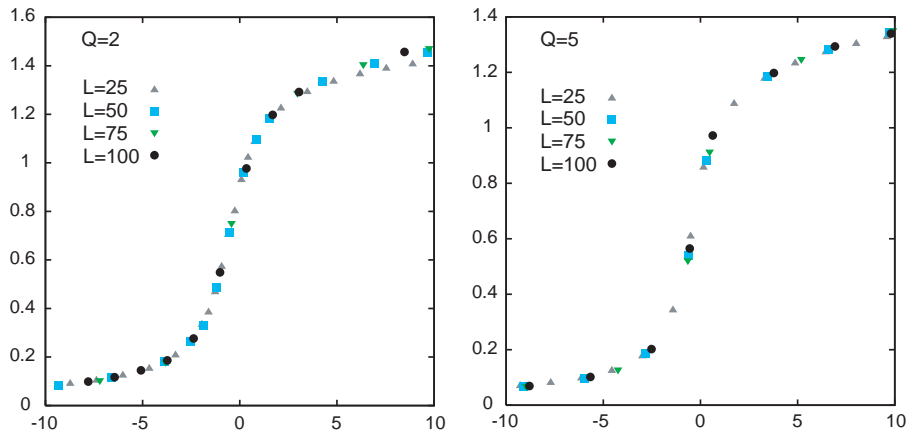}
   \end{center}
 \caption{Scaling plots of the magnetization for $Q = 2$ and $Q = 5$. $\left\langle M \right\rangle L^{1/8}$ as function of $\beta^{\ast} L$. $\beta^{\ast} = \frac{\beta - \beta^Q_{c}}{\beta^Q_c} $ is the reduced inverse temperature.}
\end{figure}

\begin{figure}[H]
   \begin{center}
    \includegraphics[width=12truecm]{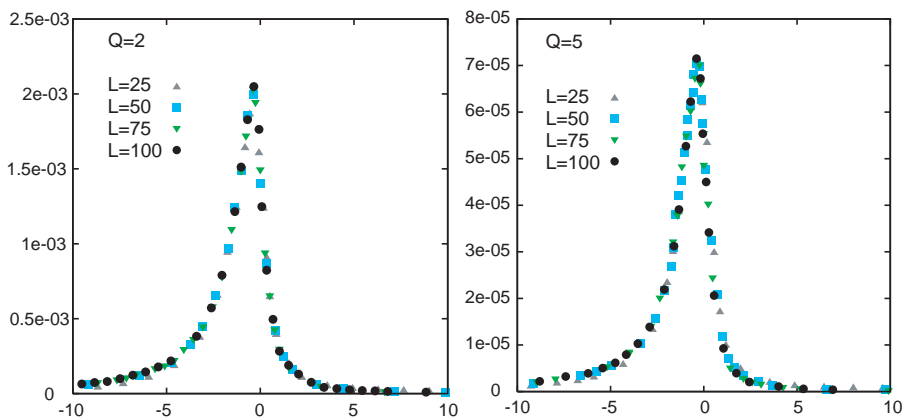}
   \end{center}
 \caption{Scaling plots of the susceptibility $\left[ M - \left\langle M \right\rangle \right]^2 \times L^{-7/4}$ as function of $\beta^{\ast} L$, for $Q = 2$ and $Q = 5$ for different system sizes.}
\end{figure}

In the following we present computed mean values of the Euler characteristic
and related Betti's numbers per site and per color for several values of $Q$ as
function of the inverse temperature $\beta$.

Periodic boundary conditions have been used which obviously lead to a
symmetry between colors $a$ and$- a$. The equilibrium configuration presents a
preference for positively or negatively oriented spins. We have chosen in the
statistics, the first one, i.e. the one corresponding to $+ Q$ boundary
conditions.

\begin{figure}[H]
   \begin{center}
    \includegraphics[width= 12truecm]{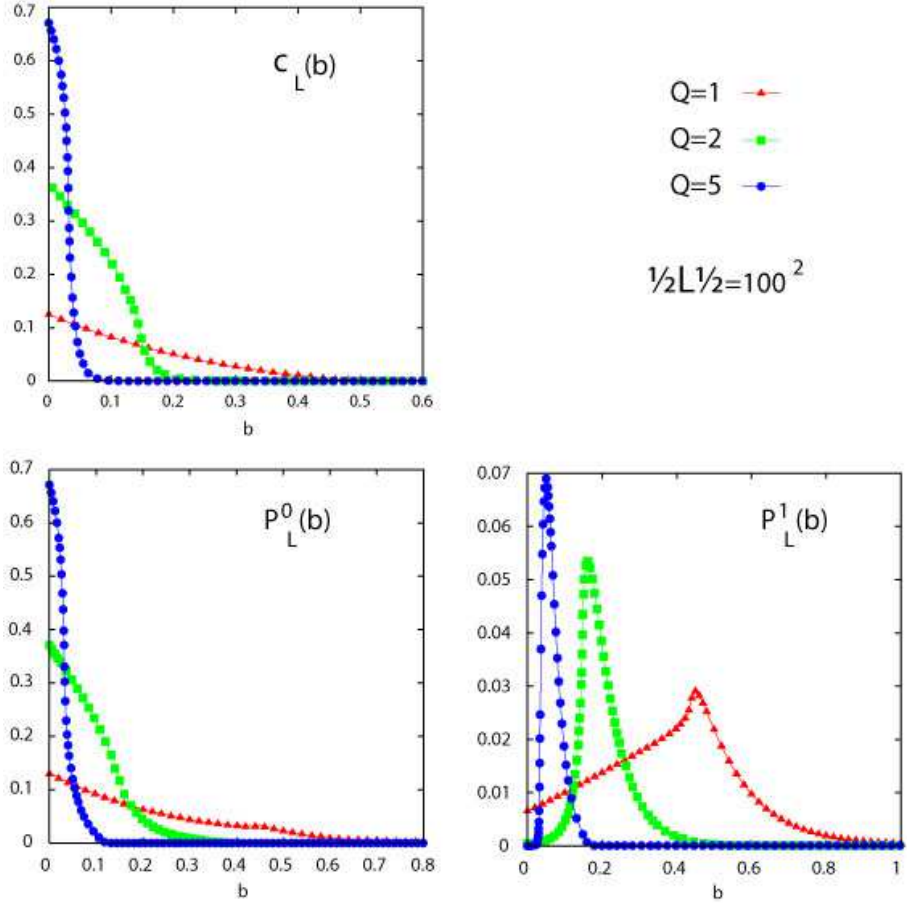}
   \end{center}
\caption{Mean values of the global Euler characteristic $\chi$ and global Betti's numbers
$\Pi_0$ and $\Pi_1$ for $Q = 1, 2, 5$ and lattice size $| \Lambda | = 100^2$.}
\label{plotglobal}
\end{figure}

Figure \ref{plotglobal} shows the behaviour of the mean values of the global
Euler characteristic $\chi^{}_{\Lambda}$ and Betti's numbers $\pi^0_{\Lambda}$
and $\pi^1_{\Lambda}$ as function of $\beta$ for $Q = 1, 2, 5$ and $| \Lambda
| = 100^2$. Recall from (\ref{EP}) that $\pi^0_{\Lambda} (\beta)$ is nothing
but the number of connected components of the complex and $\pi^1_{\Lambda}
(\beta)$ is the number of independent cycles, i.e. the number of $1
-$dimensional holes in the complex.

Seemingly, $\chi^{}_{\Lambda} (\beta)$ appears as a decreasing function of
$\beta$ and vanishes for  $\beta > \beta_c$. The decrease of
$\chi^{}_{\Lambda} (\beta)$ for $\beta < \beta_c$ is associated both to a
decrease of $\pi^0_{\Lambda} (\beta)$ and an increase of $\pi^1_{\Lambda}
(\beta)$. In addition one observes that for $\beta > \beta_c$ these two
quantities decrease and seem to compensate each other. This would explain the
vanishing of $\chi^{}_{\Lambda} (\beta)$ for $\beta \geqslant \beta_c$. As we
shall see this compensation can be understood intuitively by looking at the
behaviour of the colored Euler-Poincar\'e characteristic $\chi^{a}_{\Lambda} (\beta)$ and colored Betti's numbers $\pi^{0, a}_{\Lambda} (\beta)$,
$\pi^{1, a}_{\Lambda} (\beta)$, $ \text{$a \in S_Q$}$, $Q = 1, 2, 5$.

\begin{figure}[H]
   \begin{center}
    \includegraphics[width= 12truecm]{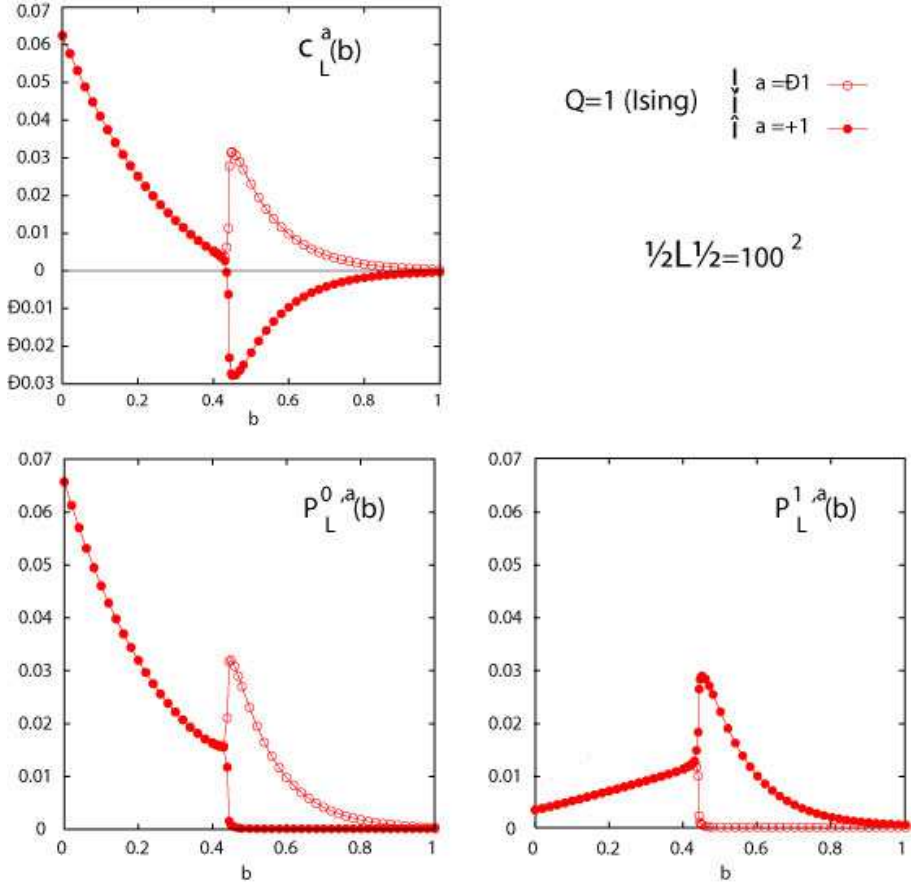}
   \end{center}
\caption{Mean
values of the colored Euler characteristic $\chi_{}^a (\beta)$ and Betti's
numbers $\pi^{0, a}_{} (\beta)$ and $\pi^{1, a}_{} (\beta)$ for $Q = 1$ and $|
\Lambda | = 100^2$.}
\label{Q1}
\end{figure}

Let us look at the Ising case $Q = 1$ for which these quantities are presented
in Figure~\ref{Q1}. We observe that for $\beta < \beta_c$ the quantities
$\chi^a_{\Lambda} (\beta)$, $\pi^{0, a}_{\Lambda} (\beta)$, $\pi^{1,
a}_{\Lambda} (\beta)$ do not depend on colors $a = \pm 1$. However they
exhibit a sharp transition in the color variable right at the critical point. 

It is also seen that above $\beta > \beta_c$, $\pi^{0, + 1}_{\Lambda} \simeq \pi^{1, - 1}_{\Lambda}\simeq 0$ and $\pi^{1, + 1}_{\Lambda} \simeq \pi^{0, - 1}_{\Lambda}$. Therefore 
$\pi^{0}_{\Lambda} \simeq \pi^{1}_{\Lambda}$ which implies that in this range of temperature the global Euler characteristic vanishes. Alternatively the behaviour of the colored Betti's numbers explain why  $\chi^{+1}_{\Lambda} (\beta) \simeq - \chi^{-1}_{\Lambda} (\beta)$ which, in view of formula (\ref{chi-additivity}), gives also that the global Euler characteristic vanishes. 
These facts can be understood intuitively at low temperatures if one has in mind the usual
picture of islands of $(-)$ inside a sea of $(+)$. Indeed, for such
configurations, say $\tmmathbf{\sigma}_{\Lambda}$, one has $\tmmathbf{\pi}^{0,
+ 1}_{\Lambda} (\tmmathbf{\sigma}_{\Lambda}) \simeq \tmmathbf{\pi}^{1, -
1}_{\Lambda} (\tmmathbf{\sigma}_{\Lambda}) \simeq 1$ and each island gives a
contribution $1$ to both $\tmmathbf{\pi}^{1, + 1}_{\Lambda}
(\tmmathbf{\sigma}_{\Lambda})$ and $\tmmathbf{\pi}^{0, - 1}_{\Lambda}
(\tmmathbf{\sigma}_{\Lambda})$.

On the other hand, when $\beta < \beta_c$ there exists only one (disordered)
phase with independent colors. This corroborates the fact that
$\chi^a_{\Lambda} (\beta)$, $\pi^{0, a}_{\Lambda} (\beta)$ and $\pi^{1,
a}_{\Lambda} (\beta)$ do not rely on colors in this domain of temperature. In
addition, the disorder increases when beta decreases, and in this domain of
temperature it is natural to expect that the number of connected components of
each color increases, while the number of independent cycles (of each color)
decreases.

\begin{figure}[H]
   \begin{center}
    \includegraphics[width= 12truecm]{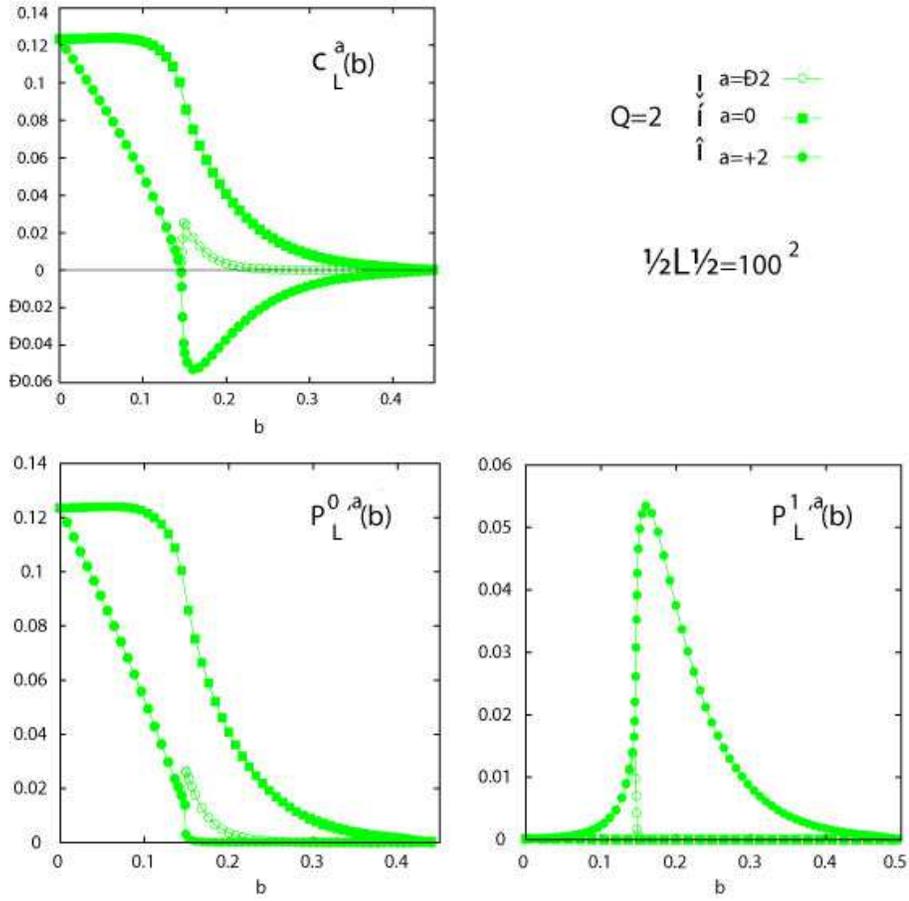}
   \end{center}
\caption{Mean
values of the colored Euler characteristic $\chi_{}^a (\beta)$ and Betti's
numbers $\pi^{0, a}_{} (\beta)$ and $\pi^{1, a}_{} (\beta)$ for $Q = 2$ and $|
\Lambda | = 100^2$.}
\label{Q2}
\end{figure}

Figure \ref{Q2} shows the results for the case $Q = 2$ ($a = 0, \pm 2$).
Considering first the colors $a = \pm 2$, one can formulate the same reasoning
as for the case $Q = 1$. The quantities of interest behave similarly as before
except for the amplitudes.

The color $a = 0$ does not present any particular feature. $\pi^{1,
0}_{\Lambda} (\beta) \simeq 0$ translates the fact that this color is
dominated by the two extremal states $a = - 2, + 2$ for all values of $\beta$.
This is reflected by noting that $\chi^0_{\Lambda} (\beta) \simeq \pi^{0,
0}_{\Lambda} (\beta)$ in Figure  \ref{Q2}.

\begin{figure}[H]
   \begin{center}
    \includegraphics[width= 12truecm]{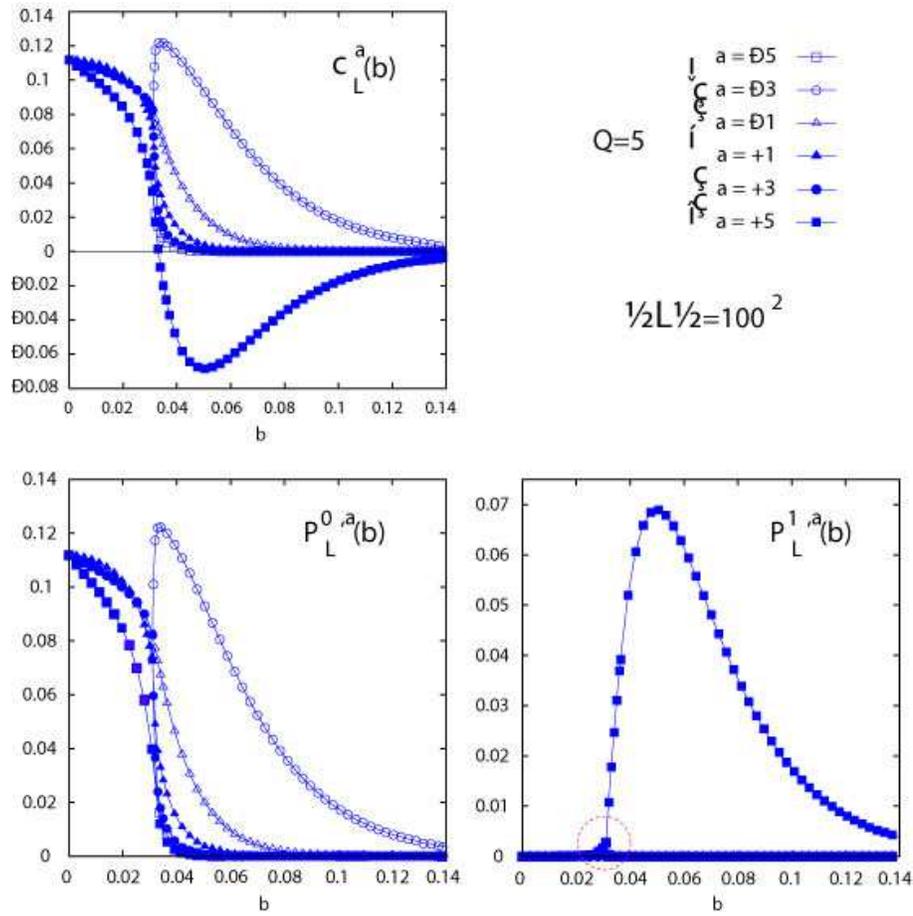}
   \end{center}
\caption{Mean values of the colored Euler characteristic $\chi_{}^a (\beta)$ and Betti's
numbers $\pi^{0, a}_{} (\beta)$ and $\pi^{1, a}_{} (\beta)$ for $Q = 5$ and $|
\Lambda | = 100^2$.}
\label{Q5}
\end{figure}

In Figure \ref{Q5} we present our results for the case $Q = 5$ ($a = \pm
5, \pm 3, \pm 1$). We observe that for $\beta > \beta_c$ the main contribution
to the number of connected components comes from color $a = - 3$.

Figure \ref{zoompi1} below gives an enlargement of Figure \ref{Q5}
showing the behavior of the other colors. Here also a sharp transition is
observed at the critical point.

\begin{figure}[H]
   \begin{center}
    \includegraphics[width= 8truecm]{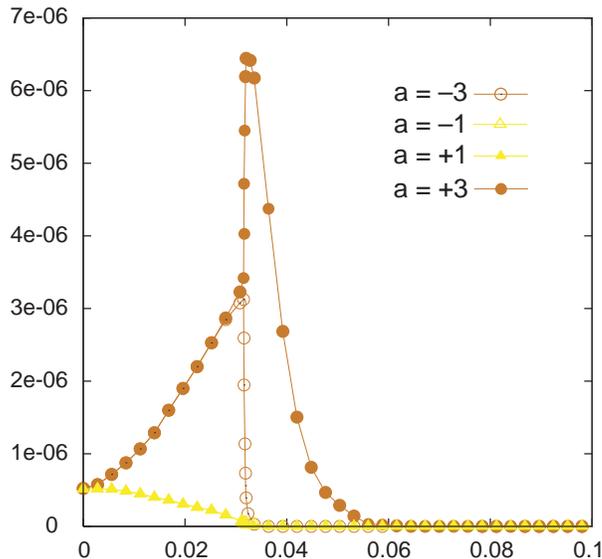}
   \end{center}
\caption{Enlargement of Figure \ref{Q5} near the critical point.}
\label{zoompi1}
\end{figure}

All numerics have been performed with careful thermalization control (up to $5.10^{6}$ MC steps before measurements). Data simulation errors were controlled through binning analysis. 
In all figures, the error bars are smaller than the sizes of the symbols. 
The Hoshen-Kopelman cluster search algorithm {\cite{HK}} for colored connected components
identification has been used.

\section{Conclusion}

In {\cite{BGRS}} we have already shown that for the Random Cluster Model, the
mean value of the Euler-Poincar\'e characteristic with respect to the
Fortuin-Kasteleyn measure exhibits a non trivial behaviour at the critical
point. Namely, there we found that, in dimension $2$, it is either zero or
discontinuous at the transition point.

Here, for the Ising model with arbitrary spins on $\mathbbm{Z}^2$, our numerics show that the
topology of configurations reveals the signature of the phase transition. In
particular this leads us to following conjecture

{\tmstrong{Conjecture}}: in the thermodynamic limit, $\chi^{}_{\Lambda
\uparrow \mathbbm{Z}^2} (\beta) > 0$ for $\beta < \beta_c$ and
$\chi^{}_{\Lambda \uparrow \mathbbm{Z}^2} (\beta) = 0$ for $\beta > \beta_c$.

We think that the vanishing of $\chi^{}_{\Lambda \uparrow \mathbbm{Z}^2}
(\beta)$ can be shown at large $\beta$ by perturbative Peierls type arguments
in view of the expression (\ref{qlocalcolor}) of the local Euler characteristic.

It would be interesting to study also these quantities in models with color
symmetry like the Potts or Clock models \cite{WP}.

{\subsection*{Acknowledgments}}

Warm hospitality and Financial support from the BiBoS research Center,
University of Bielefeld and Centre de Physique Th\'eorique, CNRS Marseille are
gratefully acknowledged. One of the authors, Ch. D., acknowledges financial
support from the IRCSET embark-initiative postdoctoral fellowship scheme. We thank the referee for constructive remarks.


\begin{thebibliography}{10}
  \bibitem[1]{Ser}Serra J, image analysis and
  mathematical morphology, 1982 \tmtextit{Academic Press}, London.
  
  \bibitem[2]{Me1}Mecke K,  morphology of spatial patterns : porous
  media, spinodal decomposition, and dissipative structures, 1997 
  \tmtextit{Acta Physica Polonica B}, \tmtextbf{28} 1747--1782.
  
  \bibitem[3]{Me2}Mecke K, exact moments of curvature measures in
  the boolean model, 2001  \tmtextit{Acta Physica Polonica B},
  \tmtextbf{102} 1343--1381.
  
  \bibitem[4]{Win}Winkler G,  image analysis, Random
  Fields and Markov Chain Monte Carlo \tmtextit{Springer}, 2003.
  
  \bibitem[5]{Wa}Wagner H, Euler characteristic for archimedean
  lattices,  2000  \tmtextit{Unpublished}.
  
  \bibitem[6]{BGRS}Blanchard Ph, Gandolfo D, Ruiz J, and Shlosman S, 
  on the Euler--Poincar\'e characteristic of the random cluster
  model, 2003  \tmtextit{Mark. Proc. Rel. Fields}, \tmtextbf{9} 523.
  
  \bibitem[7]{SyEs}Sykes M F  and Essam J W, exact critical
  percolation probabilities for site and bond problems in two dimensions, 1964
   \tmtextit{J. Math. Phys.} , \tmtextbf{5} 1117--1121.
  
  \bibitem[8]{Gri}Grimmet G, Percolation, 1999
  \tmtextit{Springer, 2nd edition}.
  
  \bibitem[9]{Gr1} Griffiths R B, rigorous results for Ising
  ferromagnet of arbitrary spins, 1969  \tmtextit{J. Math. Phys.},
  \tmtextbf{10} 1559.
  
  \bibitem[10]{Gr} Griffiths R B, correlations in Ising
  ferromagnets II., 1967  \tmtextit{J. Math. Phys.},
  \tmtextbf{8} 484--489.
  
  \bibitem[11]{BIN}Binder K, finite size scaling analysis of Ising
  model block distribution functions, 1981  \tmtextit{Z. Phys.}
  \tmtextbf{43} 119.
  
  \bibitem[12]{NB}Nicolaides D and Bruce A D, universal configurational structure in two-dimensional    
  scalar models, 1988  \tmtextit{J. Phys. A: Math. Gen.}
  \tmtextbf{21} 233--243.

  \bibitem[13]{CD}Chen X S and Dohm V, nonuniversal finite--size scaling in anisotropic systems,  
  2004  \tmtextit{Phys. Rev. E}
  \tmtextbf{70} 056136.

    \bibitem[14]{SS}Selke W and Shchur L N, critical Binder cumulant in two--dimensional anisotropic Ising models, 2005  \tmtextit{J. Phys. A: Math. Gen.}
  \tmtextbf{38} L739--L744.

  \bibitem[15]{HK}Hoshen J and Kopelman R, percolation and cluster
  distribution I., cluster multiple labeling technique and critical
  concentration algorithm, 1976  \tmtextit{Phys. Rev. B},
  \tmtextbf{14}:3438.
  
   \bibitem[16]{WP} work in progress.


\end{thebibliography}
\end{document}